\documentclass[conference]{IEEEtran}

\usepackage{amsfonts}
\usepackage{amssymb}
\usepackage{graphicx} 
\usepackage{epstopdf}
\usepackage{subfigure}
\usepackage{tikz}

\graphicspath{{./figures/}}
\DeclareGraphicsExtensions{.eps}

\ifCLASSINFOpdf
\else
\fi

\usepackage[cmex10]{amsmath}
\usepackage{array}
\usepackage{mdwmath}
\usepackage{mdwtab}
\usepackage{eqparbox}
\usepackage{fixltx2e}
\usepackage{stfloats}
\usepackage{url}

\newcommand{\SNR}{\mathsf{SNR}}
\newcommand{\Q}{\mathbb{Q}}
\newtheorem{defn}{Definition}
\newtheorem{theorem}{Theorem}

\begin{document}
%
\title{Compute-and-Forward for the Interference Channel: Diversity Precoding}

\author{\IEEEauthorblockN{Ehsan E. Khaleghi}
\IEEEauthorblockA{
Communications \& Electronics Department\\
TELECOM ParisTech\\
Paris, France\\
Email: ehsan.ebrahimi-khaleghi@telecom-paristech.fr}
\and
\IEEEauthorblockN{Jean-Claude Belfiore}
\IEEEauthorblockA{Communications \& Electronics Department\\
TELECOM ParisTech\\
Paris, France\\
Email: jean-claude.belfiore@telecom-paristech.fr}}




\maketitle

\begin{abstract}
Interference Alignment is a new solution to overcome the problem of interference in multiuser wireless communication systems. Recently, the Compute-and-Forward (CF) transform has been proposed to approximate the capacity of $K$-user Gaussian Symmetric Interference Channel and practically perform Interference Alignment in wireless networks. However, this technique shows a random behavior in the achievable sum-rate, especially at high SNR. In this work, the origin of this random behavior is analyzed and a novel precoding technique based on the Golden Ratio is proposed to scale down the fadings experiences by the achievable sum-rate at high SNR.

\end{abstract}


\vspace*{0.2cm}



\IEEEpeerreviewmaketitle

\section{Introduction}

Nowadays multiuser interference is one of the most challenging problems encountered in present wireless communication systems, particularly with growing number of subscribers as well as the decreasing size of cells for cellular systems, increasing demand in terms of transmission rates and channel limits. We should think through methods which eliminate or uses interference to recover desire information in a proper way in our communication systems. Interference Alignment (IA) is an interference management technique that achieves a linear scaling of the network throughput with the number of source-destination pairs, a scaling that would be impossible with interference avoidance. Two alignment approaches are known in literature: linear interference alignment for time-varying channels \cite{Book:Syed} and non-linear interference alignment for static single-antenna channels \cite{IEEE:Motahari}. 
 
\subsection{Related Work}

From information theoretical perspective, this issue is modeled by the interference channel introduced many years ago in \cite {IEEE:Ashlswede} and \cite {Berkeley:Shannon}. Still it remains one of the most important challenges in the domain of multiuser information theory. In two-user interference channel, a significant progress had been made for the case of \textit{strong} \cite{IEEE:Carlerial} and \textit{very strong} interference \cite{IEEE:Jafarian} channels. Indeed, it is natural to overcome the problem of achievable sum-rate described in \cite{IEEE:Gastpar}, for 2-user systems before generalizing it for $K-$user case, which ${K>2}$. 

Among existing interference management techniques, we focus in this work on IA. This novel framework will be used to design actual codes. First lattice-based codes are designed for channels with integer-valued coefficients and later extended to real-valued (resp. complex-valued) channel coefficients. In particular we are interested in lattice-based IA using the CF framework. Originally introduced by Nazer and Gastpar as relaying strategy in \cite{IEEE:Gastpar}. The CF allows relay nodes to decode and forward linear equations of originally transmitted messages using the noisy linear combinations provided by the channel. Upon receiving enough linear combinations, the destination can retrieve the original data flows with higher transmission rates compared to traditional relaying techniques. At high $\mathsf{SNR}$, the computation rate can be maximized by choosing equation coefficients close to the channel coefficients. Many works have been done to analysis the Degrees of Freedom (DoF) for CF. Among which, Nilsen \textit{et al.} have used the approach of CF to show achievability results for DOF \cite{IEEE:Niesen}. For what concerns interference management, the CF has been used by Ordentlich \textit{et al.} in \cite{IEEE:Erez} to show achievability results. The alignment problem can be formulated as that of solving an overdetermined system of equations with respect to a subset of unknowns and can be cast into the familiar language of vector spaces \cite{IEEE:Maddah-Ali}.

\subsection{Summary of Paper Results}

\begin{figure}[!h]
\centering
\includegraphics[width=2.5in]{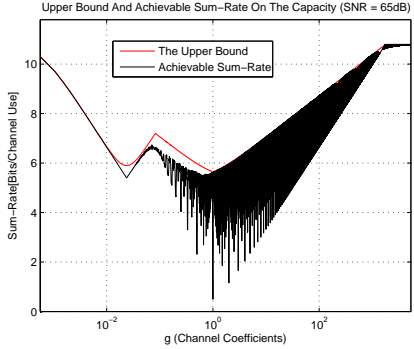}
\caption{Upper and lower bounds on the capacity of a 2-user Gaussian symmetric interference channel with respect to the cross-gain \emph{g} and the CF scheme defined in \cite{IEEE:Erez}.}
\label{fig:Ts1}
\end{figure}

Our basic strategy is to consider the computation rate, defined in \cite{IEEE:Gastpar}, for Gaussian Symmetric Interference Channels (GS-IFC) and the new scheme of CF described in \cite{IEEE:Erez}, for modeling 2-user GS-IFC, and improving its achievable sum-rate. In this paper we assume that there is no need of channel side information at transmitters. If we consider the same method used in \cite{IEEE:Erez} to simulate the achievable sum-rate for 2-user GS-IFC, we get the performance showed in Fig.~\ref{fig:Ts1}.
We are interested in the fractal behavior of the sum-rate at high values of $\mathsf{SNR}$, when using the CF transform. In this case, as it can be seen in Fig.~\ref{fig:Ts1}, the achievable sum rate suffers from deep fadings and it can change dramatically, even for a small interfering gain variation. 

In the second part of this work we will introduce the channel model and the lattice structures. In the third part, by considering the main frame work of \cite{IEEE:Erez} and \cite{IEEE:Gastpar}, we will model the achievable sum-rate for 2-user GS-IFC. After defining correspondent quadratic form, we will introduce the \lq\lq{}Golden Ratio\rq\rq{} and its equivalent structures to approximate channel coefficients. These new approximated channel coefficients will help us to avoid deep fadings and improves the achievable sum-rate by using just one time-slot to send each codewords to destination, but it will have its own disadvantages. Finaly in the last part of this work we will define our new method of CF transform to send codewords through channel to destinations in $n$ different time-slots, by using \textit{Precoders} combined with \textit{Golden Ratio} at transmitters level. This method improves significantly the final achievable sum-rate, and we can limit deep fadings of previous works.


\section{channel model and lattice structure}
\subsection{Channel Model}
 
In this paper, the channel model is the $K-$user CS-IFC. 
\begin{figure}[h]
 \centering
 \begin{tikzpicture}

      \draw (0,0) rectangle (0.7,-0.7);
      \node at (-0.75,-0.35) {$w_1$};
        \draw [->] (-0.52,-0.35)--(0,-0.35);  
      \node at (0.35,-0.35) {$\mathcal{E}_1$};
       \draw [->] (0.7,-0.35)--(2.25,-0.35); 
        \node at (0.4,-0.95) {$\mathrm{TX}_1$}; 
       \node at (0.9,-0.15) {${x}_1$};
         \draw (2.4,-0.35) circle [radius=0.15];
          \draw  (2.3,-0.35)--(2.5,-0.35);
          \draw (2.4,-0.25)--(2.4,-0.45);
          \node at (1.8,-0.15) {$1$};
              \node at (2.45,0.25) {${z}_1$};  
          \draw [->] (2.4,0.18)--(2.4,-0.2); 
          \draw [->] (2.55,-0.35)--(3.25,-0.35); 
           \node at (3,-0.15) {${y}_1$};
          \draw (3.25,0) rectangle (3.95,-0.7);
           \node at (3.6,-0.35) {$\mathcal{D}_1$};
             \node at (3.65,-0.95) {$\mathrm{RX}_1$}; 
             \draw [->] (3.95,-0.35)--(4.47,-0.35); 
                \node at (4.7,-0.3) {$\hat{w}_1$};

         \draw (0,-2) rectangle (0.7,-2.7);
      \node at (-0.75,-2.35) {${w}_2$};
        \draw [->] (-0.52,-2.35)--(0,-2.35);  
      \node at (0.35,-2.35) {$\mathcal{E}_2$};
       \draw [->] (0.7,-2.35)--(2.25,-2.35); 
        \node at (0.4,-2.95) {$\mathrm{TX}_2$}; 
       \node at (0.9,-2.15) {${x}_2$};
         \draw (2.4,-2.35) circle [radius=0.15];
          \draw  (2.3,-2.35)--(2.5,-2.35);
          \draw (2.4,-2.25)--(2.4,-2.45);
          \node at (1.8,-2.15) {$1$};
              \node at (2.45,-1.75) {${z}_2$};  
          \draw [->] (2.4,-1.82)--(2.4,-2.2); 
          \draw [->] (2.55,-2.35)--(3.25,-2.35); 
           \node at (3,-2.15) {${y}_2$};
          \draw (3.25,-2) rectangle (3.95,-2.7);
           \node at (3.6,-2.35) {$\mathcal{D}_2$};
             \node at (3.65,-2.95) {$\mathrm{RX}_2$}; 
             \draw [->] (3.95,-2.35)--(4.47,-2.35); 
                \node at (4.7,-2.3) {$\hat{w}_2$};

                \draw [->] (1,-0.35)--(2.3,-2.25); 
                   \draw [->] (1,-2.35)--(2.3,-0.45); 
                   \node at (1.27,-1) {$g$}; 
                    \node at (1.3,-1.65) {$g$};

 \end{tikzpicture}
 \caption{2-User Gaussian Symmetric Interference Channel.}\label{Fig:GS-IFC}
 \end{figure}
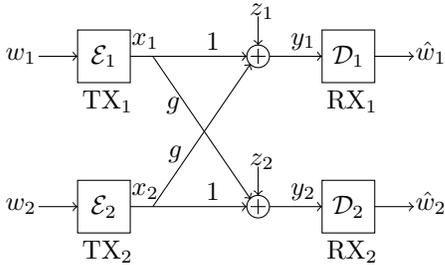

By using a simple lattice interference alignment \cite{IEEE:Erez}, the symmetric $K-$user case is approximately equivalent to the symmetric $2-$user case which is shown in Fig. \ref{Fig:GS-IFC}. This means that:
\begin{equation}
\mathbf{H}=\left[\begin{array}{cc}
1 & g\\
g & 1
\end{array}\right]
\end{equation}
The received signal is expressed by:

\begin{equation}
\label{Channel_model}
\mathbf{y} = \left[\begin{array}{c}
y_1\\
y_2
\end{array}\right] = \mathbf{H} \mathbf{x} + \mathbf{z}
\end{equation}

Where $\mathbf{x}$ denotes the input-vector, $\mathbf{y}$ the output-vector and $\mathbf{z}$ the noise-vector, all of size $K$. The components of $\mathbf{z}$ are independent Gaussian real zero-mean random variables with power equal to $\sigma^2$. The $\SNR$ is defined as $\SNR=\frac{P}{\sigma^2}$. Each transmitter satisfies the power constraint, which for \emph{n} channel uses for user \emph{i} is given by: 

\begin{equation}
\label{Channel_power_contrainte}
{\frac{1}{n}} \sum_{j=0}^{n} \vert {\mathbf{x}_\mathrm{\emph{i}}}^2 \vert \leq P_\mathrm{\emph{i}}
\end{equation}

We assume in this paper that all users have the same power constraint i.e., $P_\mathrm{i} = P$. The channel is symmetric in the third part of this paper, but in the last part of this paper when we introduce our precoding scheme, we are transforming the Symmetric Channel (SC) to Asymmetric Channel (AC). A channel is said symmetric when $H(i,j) = g$ for all $i \not= j$, and after normalization $H(i,i) = 1$ for all \emph{i}'s.

\subsection{Lattice structure}

In this paper we will use the Nested lattice framework proposed by \cite{IEEE:Erez_lattice}. This choice allows to achieve the computation rate that will be used below. A lattice $\Lambda$ is a discrete additive subgroup of ${\mathbb{R}}^n$, i.e., $\forall{t_\mathrm{\emph{1}}}, {t_\mathrm{\emph{2}}}\in\Lambda$, where ${t_\mathrm{\emph{1}}}+{t_\mathrm{\emph{2}}}\in\Lambda$ and $-{t_\mathrm{\emph{1}}},-{t_\mathrm{\emph{2}}}\in\Lambda$. Any lattice $\Lambda$ in ${\mathbb{R}}^n$ can be characterized by a ${n}\times{n}$ symmetric definite positive matrix \emph{G} called Gram matrix  or by using a ${n}\times{n}$ matrix \emph{M} called generator matrix such that:

\begin{equation}
\label{Lattice_Model}
\Lambda =  \{A=M\cdot Z: Z\in{\mathbb{Z}}^{n}\}
\end{equation}
By applying the Cholesky decomposition to matrix \emph{G} we can create an upper triangular matrix \emph{B} of size ${n}\times{n}$ which is a generator matrix of $\Lambda$. The columns of matrix \emph{B} are basis of $\Lambda$. We get: 

\begin{equation}
\label{Lattice_Gram_Base}
B=\mathrm{Cholesky}(G), \hspace*{3pt}G=B^TB
\end{equation}

A lattice $\Lambda$ is full rank if its Gram matrix is full-rank.

\section{Diophantine approximation for the 2-user GS-IFC and the Golden ratio}

We use the so-called CF Transform introduced in \cite{IEEE:Erez}. 
In \cite[Theorem 2]{IEEE:Gastpar}, the computation rate which is the maximal rate at which users can transmit codewords to destinations, when we are interested in decoding reliably the equation $\sum_i{a_i x_i}$ is given by: 

\begin{equation}
\label{Computation_rate}
R(\mathbf{h},\mathbf{a})=\frac{1}{2}\log_2^{+}\left\{\left({\parallel \mathbf{a}\parallel^{2}}-\frac{\mathsf{SNR}(\mathbf{h}^{T}\mathbf{a})^{2}}{1+\mathsf{SNR}{\parallel \mathbf{h}\parallel^{2}}}\right)^{-1}\right\}
\end{equation}

Where $\log_{2}^{+}(x)\triangleq \max(\log_{2}(x),0)$, $\mathbf{a}$ is a vector of integers of length \emph{n} which will characterize the equation we want to decode and $\mathbf{h}$ is the vector of channel coefficients. 
In this contribution, we are interested to improve the behavior of CF in the \textit{strong} and \textit{very strong} interference regimes. More precisely, we want to limit the deep fading behavior observed in Fig. \ref{fig:Ts1} for the achievable sum-rates. In the $2-$user GS-IFC for each user, $\mathbf{h}=[1,g]$ and the Interference-to-Noise Ratio (INR) is $\mathsf{INR}\triangleq g^{2}\mathsf{SNR}$. Here, $g\in\mathbb{R}$ is the channel interferer coefficient, the direct channel coefficient is normalized to be 1 and $\mathbf{a}=[x,y]$. The computation rate expressed in (\ref{Computation_rate}) can be written in this way:

\begin{equation}
\label{Computation_rate_form}
R(\mathbf{h},\mathbf{a})=\frac{1}{2}\log_{2}^{+}\left\{ \frac{\left(\frac{1}{\mathsf{SNR}}+(1+g^{2})\right)}{q\left(x,y\right)}\right\} 
\end{equation}
Where $q(x,y)$ is a definite positive quadratic form equal to:

\begin{equation}
\label{quadratic_form}
q\left(x,y\right)=(xg-y)^{2}+\frac{1}{\mathsf{SNR}}(x^{2}+y^{2}), x,y\in\mathbb{Z}
\end{equation}
From Equation (\ref{quadratic_form}), the Gram matrix can be found as: 

\[G= \left(\begin{array}{ c c } g^{2}+\frac{1}{\mathsf{SNR}} & -g \\ -g & 1+\frac{1}{\mathsf{SNR}} \end{array} \right)\] 
As a definite positive integral quadratic form, $q(x,y)$ defines a rank $2$ lattice, $\Lambda_{\mathrm{CF}}$. 

Following the method described in \cite{IEEE:Erez}, we aim to find the two successive minima $\lambda_{\mathrm{\emph{1}}}$ and $\lambda_\mathrm{\emph{2}}$ of (\ref{quadratic_form}). As the integral quadratic form is of dimension $2$, an algorithm for optimally finding the two successive minima is the Gauss reduction algorithm \cite{gauss_reduction}(which has been generalized to the LLL reduction algorithm \cite{lll_reduction} in higher dimension). Let's define matrix $B=\mathbf{Cholesky}(G)$ be a basis of $\Lambda_{\mathrm{CF}}$ and $B_{\mathrm{red}}$ be the reduced basis after Gauss reduction. $U$ is the unimodular basis change matrix. Call $G_{\mathrm{red}}=B_{\mathrm{red}}^T B_{\mathrm{red}}$ the reduced Gram matrix, then the two successive minima are the diagonal entries of $G_{\mathrm{red}}$. In this part of our work, we analyze the achievable sum-rate for different values of \emph{g}.  

The quadratic form of Equation (\ref{quadratic_form}) can be decomposed into two terms. 
\begin{itemize}
 \item First term $(xg-y)^2$, where $x$ and $y$ are integers, is obviously related to the quality of the Diophantine approximation of the real number $g$. 
 \item Second term, $\frac{1}{\SNR} (x^2+y^2)$ is a penalty that disadvantages large values of $x$ and $y$. 
\end{itemize}
We perform here an asymptotic analysis of (\ref{quadratic_form}) for which the Diophantine approximation term is the most important term. 
Values of $g$ giving a high sum-rate, are those values for which the value of the second minimum $\lambda_2$ is low. But, as the product of $\lambda_1 \lambda_2$ is a constant \cite{IEEE:Erez}, those values of $g$ are those for which the value of the first minimum $\lambda_1$ is high. 
In a high $\SNR$ analysis, this means that the real number $g$ must be hardly approximable by a rational number. 
In \cite[Chap. 2]{cassels}, this problem is considered and the Golden ratio is shown to be the most hardly approximable number (which is intuitive since its continuous fraction development gives only 1's). 

\subsection{Equivalent Numbers and Diophantine approximation}

The Golden ratio is the most hardly approximable real number. Following \cite{cassels}, we define, for a real number $\theta$, and an integer $q$, 
\begin{equation}
 \left\Vert q\theta\right\Vert=\min_{p\in \mathbb{Z}}|q\theta-p|
\end{equation}
Then, for any $\theta$, we get
\begin{equation}
 q \left\Vert q\theta\right\Vert < 5^{-\frac{1}{2}}
\end{equation}

\begin{defn}
Let $\phi=\frac{1+\sqrt{5}}{2}$ be the Golden ratio. The conjugate of $\phi$ will define as $\bar\phi=\frac{1-\sqrt{5}}{2}$. A number $g$ is equivalent to $\phi$ if:
\begin{equation}
\label{Golden_Approximation}
g=\frac{a\phi+b}{c\phi+d}\,,\;(a, b, c, d)\in\mathbb{Z\,\mathrm{~and}\,\det\left[\begin{array}{lc}
\mathrm{a} & \mathrm{b}\\
\mathrm{c} & \mathrm{d}
\end{array}\right]}=\pm1 
\end{equation}
\end{defn}

We are ready to state one of the main results \cite[Chap. I, Theorem V]{cassels}. 
\begin{theorem}\label{theor:bound}
 Let $\theta$ be irrational. Then there are infinitely many $q$ such that $q \left\Vert q\theta\right\Vert < 5^{-\frac{1}{2}}$. If $\theta$ is equivalent to $\phi$, then the constant $5^{-\frac{1}{2}}$ cannot be replaced by any smaller number. 
\end{theorem}

This result means that, if $\SNR$ is large enough, and if $g$ is a number equivalent to the Golden Ratio, then the second minimum of $q(x,y)$ is small, and so, the sum-rate of the interference channel is high. 

Now we will give an example to show that this method will give an achievable sum-rate close to the outer Bound, with a small gap between them. 


\begin{figure}
\centering
\includegraphics[width=2.5in]{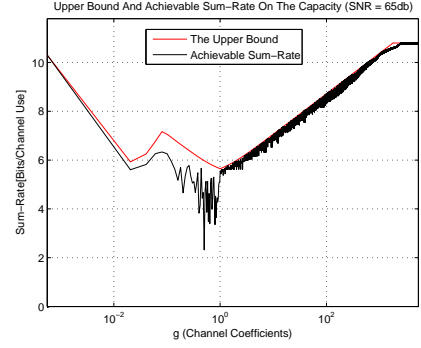}
\caption{Upper and lower bounds on the capacity of 2-user Gaussian symmetric interference channel with respect to the use of perfect approximated cross-gain \emph{$g^{'}$} equivalent number to golden ratio.}
\label{fig_approx}
\end{figure}

Suppose now the interfering coefficient $g$ is equal to the Golden Ratio, i.e. $g=\frac{1+\sqrt{5}}{2}$, When ${\mathsf{SNR}}$ is high enough, then the $2-$dimensional vector of integers which minimizes the quadratic form $q(x, y)$ defined in (\ref{quadratic_form}) will have a length equal to the first minimum. According to Theorem \ref{theor:bound}, this first minimum should satisfy $x^{2} (xg-y)^2\approx\frac{1}{5}$. So we get

\begin{equation}
\label{Lambda_1}
\lambda_\mathrm{\emph{1}}\approx\frac{1}{5x^{2}}+\mathsf{SNR}^{-1}(x^{2}+y^{2})
\end{equation}

or

\begin{equation}
\label{Lambda_1_bis}
\lambda_\mathrm{\emph{1}}\approx\frac{1}{5x^{2}}+\mathsf{SNR}^{-1}x^{2}(1+g^{2})
\end{equation}

Now, the minimum is achieved when

\begin{equation}
\frac{\partial\lambda_\mathrm{\emph{1}}}{\partial x} = -\frac{2}{5x^{3}}+2x\mathsf{SNR}^{-1}(1+g^{2})=0
\end{equation}

giving

\begin{equation}
\label{x_optimal}
x_\mathrm{opt}=\sqrt[4]{\frac{\mathsf{SNR}}{(5(1+g^{2}))}}
\end{equation}

and

\begin{equation}
\label{Lambda_1_bis_1}
\lambda_\mathrm{\emph{1}}\approx2\sqrt{\frac{1+g^2}{5\mathsf{SNR}}}
\end{equation}

As $\lambda_\mathrm{\emph{1}}\lambda_\mathrm{\emph{2}}\approx(1+g^{2})\mathsf{SNR}^{-1}$ \cite{IEEE:Erez}, we will get 

\begin{equation}
\label{Lambda_2}
\lambda_\mathrm{\emph{2}}\approx\sqrt{\frac{5}{4}\mathsf{SNR}^{-1}(1+g^2)}
\end{equation}

So the final rate is

\begin{equation}
\label{Example_final_rate}
\begin{array}{lc}
R\approx\frac{1}{4}\log_{2}^{+}(\mathsf{SNR}(1+g^2))-\frac{1}{4}\log_{2}^{+}(\frac{5}{4})\\
\approx{R_\mathrm{Up.Bound}}-0.08
\end{array}
\end{equation}
 
Where $R_\mathrm{Up.Bound}$ is the Upper Bound of the rate for a $\mathsf{SNR}$ sufficiently large. So the gap between Upper Bound and achievable sum-rate is very small. 

Another illustration of this result which states that, at high $\SNR$, channel coefficients equivalent to the Golden ratio gives a sum-rate close to the Upper Bound, consists of plotting the same curve as in Fig. \ref{fig:Ts1}, but only sampling those $g$'s which correspond to numbers equivalent to the Golden ratio. 

In this method, we choose to sample $g$ at values of the form 

\begin{equation}
\label{g_form}
 g=\frac{a\phi+b}{c\phi+d} \nonumber
\end{equation}

Where $a,b,c,d\in\mathbb{Z}$, are not too big. Fig.~\ref{fig_approx} shows that the Upper Bound is almost achievable without any fading behavior for the strong and very strong interference regimes. The other regimes remain untouched. For some specific values of $g$ this method is not suitable, because the approximation error between channel coefficient and its equivalent to the Golden ratio will not be negligible. If we consider this approximation error in the achievable sum-rate, the fractal behavior in terms of deep fadings will appear again. We need to use a method which could hold up for any values of channel coefficient $g$. In the next section, we will introduce a new method to improve the behavior of achievable sum-rate for any value of $g$. 

\section{Compute-and-forward transform with $n$ time slots}

For large values of $\SNR$, the best choices for $g$ are numbers equivalent to the Golden Ratio. The worst choices are rational numbers as it is 
shown below.  
Suppose that $g=\frac{p}{q}\in\Q$. In this case, the minimum of $q(x,y)$
(for a sufficiently high $\SNR$) is given by setting $x=q,y=p$
which gives $\lambda_{1}=\left(p^{2}+q^{2}\right)\SNR^{-1}$. As 

\begin{equation}
\frac{\left(q^{2}\SNR^{-1}+p^{2}+q^{2}\right)^{2}}{\lambda_{1}\lambda_{2}}=q^{2}\left(p^{2}+q^{2}\right)\SNR \nonumber
\end{equation}

we get 

\begin{eqnarray*}
\lambda_{1}\lambda_{2} & = & \frac{\left(\SNR^{-1}+1+g^{2}\right)^{2}}{\left(1+g^{2}\right)\SNR}\\
 & \approx & \left(1+g^{2}\right)\SNR^{-1}
\end{eqnarray*}

which gives $\lim_{\SNR\rightarrow\infty}\lambda_{2}=\frac{1}{q^{2}}$ and 
\begin{equation}
\lim_{\SNR\rightarrow\infty}R=\frac{1}{2}\log_{2}\left(p^{2}+q^{2}\right)\nonumber
\end{equation}

So the rate does not scale with $\SNR$ and rational numbers will mainly be responsible of deep fadings at high $\SNR$. The choice of $p$ and $q$ are important, we must choose them in a way to have $p$ and $q$ as smallest possible integer.


\begin{figure}[!h]
\centering
\includegraphics[width=2.5in]{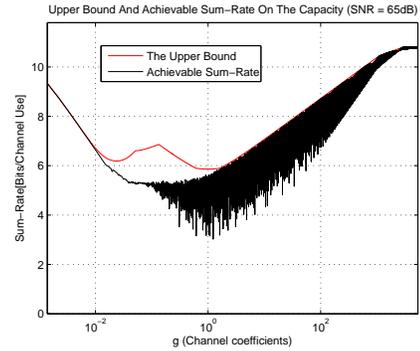}
\caption{Upper bound and achievable rate versus $g$ for a $2-$user Gaussian symmetric interference channel for $2$ time-slots.}
\label{Fig_65db_Ehsan_and_Jean-Claude}
\end{figure}

In general, transmitters want to send their own codewords to the destinations but the inference will play an important role on the performance of achievable sum-rates. To avoid and limit deep fadings in interested regimes, we have decided to send the codewords to destination by using $n$ different time-slots. Our idea now is to precode for each time slot the transmitted codewords by multiplying them, at the transmitters, by a real number $\eta$. Of course, there will always be values of $g$ such that $\eta\cdot g$ is rational (the worst case). But, by using at time slot $i$, a value of $\eta_i$ different 
of $\eta_j$ for $j\neq i$ such that, if $\eta_i\cdot g \in\Q$, then, $\eta_j\cdot g \notin\Q,\forall j\neq i$. By using this strategy, only 
one time slot over $n$ will result in a small sum-rate, for all values of $g$. 

By using this proposed scheme the 2-user GS-IFC will transform to 2-user Gaussian Asymmetric Interference Channel (GA-IFC). The new GA-IFC is shown in Fig. \ref{Fig:GA-IFC}. 

For each user the new channel coefficients vectors are respectively: $h_\mathrm{Ts=i}=[1, \eta_i g]$ for $i=1,\ldots,n$.
The quadratic form, at time slot $i$ is:

\begin{equation}
\label{Quadratic_ts_i_bis} 
q_{Ts=i}\rq\left(x,y\right)=(xg\eta_i-y)^{2}+\frac{1}{\mathsf{SNR}}(x^{2}+y^{2})
\end{equation}
 
More precisely, the Gram matrix corresponding to  (\ref{Quadratic_ts_i_bis}),  will be $G_{i}$, 

\[G_i= \left(\begin{array}{ c c } (\eta_i g)^{2}+\frac{1}{\mathsf{SNR}} & -g\eta_i \\ -g\eta_i & (1+\frac{1}{\mathsf{SNR}}) \end{array} \right)\]
In this proposed scheme, achievable sum-rate will be calculated for each time-slot separately. The final achievable sum-rate, at the end of the last time-slot, will be the average value of the sum rates over the time-slots. 

\begin{figure}[h]
 \centering
 \begin{tikzpicture}

      \draw (0,0) rectangle (0.7,-0.7);
      \node at (-0.85,-0.35) {$w_{i,1}$};
        \draw [->] (-0.52,-0.35)--(0,-0.35);  
      \node at (0.35,-0.35) {$\mathcal{E}_1$};
       \draw [->] (0.7,-0.35)--(2.25,-0.35); 
        \node at (0.4,-0.95) {$\mathrm{TX}_1$}; 
       \node at (1,-0.15) {${x}_{i,1}$};
         \draw (2.4,-0.35) circle [radius=0.15];
          \draw  (2.3,-0.35)--(2.5,-0.35);
          \draw (2.4,-0.25)--(2.4,-0.45);
          \node at (1.8,-0.15) {$1$};
              \node at (2.45,0.3) {${z}_{i,1}$};  
          \draw [->] (2.4,0.18)--(2.4,-0.2); 
          \draw [->] (2.55,-0.35)--(3.25,-0.35); 
           \node at (3,-0.15) {${y}_{i,1}$};
          \draw (3.25,0) rectangle (3.95,-0.7);
           \node at (3.6,-0.35) {$\mathcal{D}_1$};
             \node at (3.65,-0.95) {$\mathrm{RX}_1$}; 
             \draw [->] (3.95,-0.35)--(4.47,-0.35); 
                \node at (4.9,-0.3) {$\hat{w}_{i,1}$};

         \draw (0,-2) rectangle (0.7,-2.7);
      \node at (-0.85,-2.35) {${w}_{i,2}$};
        \draw [->] (-0.52,-2.35)--(0,-2.35);  
      \node at (0.35,-2.35) {$\mathcal{E}_2$};
       \draw [->] (0.7,-2.35)--(2.25,-2.35); 
        \node at (0.4,-2.95) {$\mathrm{TX}_2$}; 
       \node at (1,-2.55) {${x}_{i,2}$};
         \draw (2.4,-2.35) circle [radius=0.15];
          \draw  (2.3,-2.35)--(2.5,-2.35);
          \draw (2.4,-2.25)--(2.4,-2.45);
          \node at (1.8,-2.15) {$\eta_i$};
              \node at (2.45,-1.70) {${z}_{i,2}$};  
          \draw [->] (2.4,-1.82)--(2.4,-2.2); 
          \draw [->] (2.55,-2.35)--(3.25,-2.35); 
           \node at (3,-2.15) {${y}_{i,2}$};
          \draw (3.25,-2) rectangle (3.95,-2.7);
           \node at (3.6,-2.35) {$\mathcal{D}_2$};
             \node at (3.65,-2.95) {$\mathrm{RX}_2$}; 
             \draw [->] (3.95,-2.35)--(4.47,-2.35); 
                \node at (4.9,-2.3) {$\hat{w}_{i,2}$};

                \draw [->] (1,-0.35)--(2.3,-2.25); 
                   \draw [->] (1,-2.35)--(2.3,-0.45); 
                   \node at (1.27,-1) {$g$}; 
                    \node at (1.3,-1.65) {$\eta_i\cdot g$};

 \end{tikzpicture}
 \caption{2-User Gaussian Asymmetric Interference Channel for Ts = $i$.}\label{Fig:GA-IFC}
 \end{figure}

For comparing the performance of our proposed strategy and the strategy used in \cite{IEEE:Erez}, we choose $\mathsf{SNR}=65dB$. First we decide to send codewords to destinations by using just two different time-slots. For receiver 1 (RX1) in each time-slot the channel coefficients vectors are respectively: $h_{Ts=1} = [1,\phi g]$, and  $h_{Ts=2} = [1,\bar\phi g]$. At the end of the second time-slot, the new achievable sum-rate expression and quadratic forms to minimize are:

\begin{equation}
\label{R_tot_ts_2}
\begin{array}{lc}
R_{Final}\rq=\frac{1}{2}\left\{R_{Ts=1}\rq+R_{Ts=2}\rq\right\} 
\end{array}
\end{equation} 

With

\begin{equation}
\label{R_Ts_1}
 R_{Ts=1}\rq=\frac{1}{2}\log_{2}^{+}\left\{ \frac{(\frac{1}{\mathsf{SNR}}+(1+(\varphi g)^{2}))}{(q_{Ts=1}\rq\left(x,y\right))}\right\}  
\end{equation}
 
\begin{equation}
\label{R_Ts_2}
 R_{Ts=2}\rq=\frac{1}{2}\log_{2}^{+}\left\{ \frac{(\frac{1}{\mathsf{SNR}}+(1+(\bar\varphi g)^{2}))}{(q_{Ts=2}\rq\left(x\rq,y\rq\right))}\right\}  
\end{equation}

 And

\begin{equation}
\label{Quadratic_ts_1_bis} 
q_{Ts=1}\rq\left(x,y\right)=(xg\varphi-y)^{2}+\frac{1}{\mathsf{SNR}}(x^{2}+y^{2}), x,y\in\mathbb{Z}
\end{equation}
 
\begin{equation}
\label{Quadratic_ts_2_bis} 
q_{Ts=2}\rq\left(x\rq,y\rq\right)=(x\rq g\bar\varphi-y\rq)^{2}+\frac{1}{\mathsf{SNR}}(x\rq^{2}+y\rq^{2}), x\rq{},y\rq{}\in\mathbb{Z}
\end{equation}

Indeed, with ${q\rq{}}_ {Ts=1}$ and ${q\rq{}}_{Ts=2}$ we can create the two Gram matrices $G_{1}$ and $G_{2}$ corresponding to the two positive new quadratic forms. With these two Gram matrices we can use the \emph{single lattice codes} and \emph{lattice Han-Kobayashi} described in \cite{IEEE:Erez} for different interference regimes to find the achievable sum-rate. The two Gram matrices are modeled as: 

\[G_1= \left(\begin{array}{ c c } ((\varphi g)^{2}+\frac{1}{\mathsf{SNR}}) & -g\varphi \\ -g\varphi & (1+\frac{1}{\mathsf{SNR}}) \end{array} \right)\]

And

\[G_2= \left(\begin{array}{ c c } ((\bar\varphi g)^{2}+\frac{1}{\mathsf{SNR}}) & -g{\bar{\varphi}} \\ -g{\bar{\varphi}} & (1+\frac{1}{\mathsf{SNR}}) \end{array} \right)\]

By using precoders, the two $\mathsf{INR}$s will be different in each time-slot; we define these two $\mathsf{INR}$s, such as:

\begin{equation}
\begin{array}{lc}
\label{INRs_ts}
In~Ts=1 : \mathsf{INR}_{1} \triangleq ({\varphi}g)^{2}\mathsf{SNR} \\
In~Ts=2 : \mathsf{INR}_{2} \triangleq (\bar{\varphi}g)^{2}\mathsf{SNR}
\end{array}
\end{equation}

The expression of Upper Bound defined in \cite{IEEE:Gap_Etkin} must be adapted for our proposed scheme. In this case for two time-slots the Upper Bound will be:

\begin{equation}
\label{Up_Bound}
R_{U.B,Final} = \frac{1}{2}(R_{U.B, Ts=1}+R_{U.B, Ts=2})
\end{equation}


\begin{figure}[!h]
\centering
\includegraphics[width=2.5in]{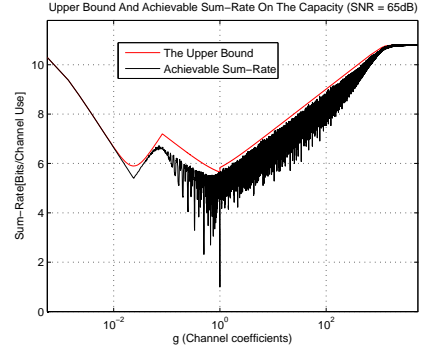}
\caption{Upper bound and achievable rate versus $g$ for a $2-$user Gaussian symmetric interference channel for $7$ time-slots.}
\label{fig:Ts7}
\end{figure}

In Fig. \ref{fig:Ts1} we can see the performance of method described in \cite{IEEE:Erez} for $\mathsf{SNR}=65dB$. Fig. \ref{Fig_65db_Ehsan_and_Jean-Claude} is the performance of our proposed scheme for the same value of $\mathsf{SNR}$ just by using $2$ time-slots. 


\begin{figure}[!h]
\centering
\includegraphics[width=2.5in]{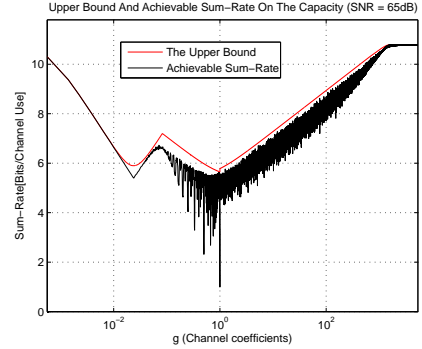}
\caption{Upper bound and achievable rate versus $g$ for a $2-$user Gaussian symmetric interference channel for $13$ time-slots.}
\label{fig:Ts13}
\end{figure}

As it can be seen in Fig. \ref{fig:Ts1} for strong and very strong interference regimes, we have deep fadings. For some channel coefficients, we can have a maximum gap of order $2.5~[Bits/Channel~Use]$. After using precoders and $2$ time-slots this gap could be reduce to $1.2~[Bits/Channel~Use]$, this is the benefit of using two time-slots.
By increasing the number of time-slots we can limit more fadings. Figure \ref{fig:Ts7} and \ref{fig:Ts13} shows the achievable sum-rate when using respectively $7$ and $13$ time-slots.

For weak and intermediate interference regimes, we have decided to send codewords to destinations by using just one time-slot. This strategy will increase the achievable sum-rate by using Han-and-Kobayashi method. But for strong and very strong interference regimes, 7 and 13 time-slots were used. The $\eta_{i}$ are all equivalent to the Golden ratio slightly greater than 1, in consequence this choice will help us to keep the Upper Bound in its original form and the achievable sum-rate will be higher. We can assume that our proposed scheme have limited the deep fadings and improved the achievable sum-rate. 

In a future work, we are going to evaluate the influence of using different time-slots with specific precoders for each time-slot. We will try to define the optimum number of time-slots and precoder coefficients for the case of 2-user GS-IFC to eliminate deep fadings in all regimes.

\section{Conclusion}
In this work, based on, the main frame work of \cite{IEEE:Erez} and \cite{IEEE:Gastpar}, we have developed two different schemes for 2-user GS-IFC. First, we have characterized what are the best (equivalent to the Golden ratio) and the worst (rational) channels. 
In order to avoid the worst-case channels, we have proposed to use a precoder (independent of the channel values and not using any channel side information at the transmitters) for sending codewords to destinations in different time-slots. 
The proposed scheme has shown an important reduction of the fading behavior of the sum-rate, similar to what is obtained in fast fading channels when a diversity technique is used. 
Many things remain to do, among which, 
\begin{itemize}
 \item The medium interference regimes. 
 \item The optimal number of Ts to be used. 
 \item Generalization to the $K-$user asymmetric interference channel. 
\end{itemize}

\end{document}